# Impacts of Histories and Models on LLM Grading: A Study in Advanced Software Engineering Courses

Qilin Zhou, Zhuo Wang, Yue Li, and W.K. Chan*
{qilin.zhou, zhuo.wang, yueli345-c}@my.cityu.edu.hk, wkchan@cityu.edu.hk
City University of Hong Kong, Hong Kong SAR, China
*corresponding author

***Abstract*—Graduate-level research reading report assessment creates a substantial labor burden for educators. While large language models (LLMs) hold great potential for automating academic grading, their reliability for this specialized task remains understudied, particularly regarding grading consistency, the lack of which represents a primary obstacle to educational fairness. This paper proposes a human-aligned LLM-assisted grading workflow and presents a case study based on 180 student submissions from a graduate advanced software engineering course. We evaluate two mainstream LLMs, Grok and GPT, in terms of grading consistency and alignment with human scores. We find LLMs exhibit distinct levels of intra-model consistency and significant inter-model grading inconsistencies, while simple ensemble approaches cannot improve alignment with human evaluation. Critically, continuous interaction history drives systematic drift in models' grading standards away from human expert scores. Our findings demonstrate LLMs' potential in reducing grading workload for educators in graduate edu-cation, while highlighting that indiscriminate LLM grading may introduce systemic unfairness, suggesting that specific operational practices are required to mitigate such disparities.**

## I. Introduction

The emergence of Large Language Models (LLMs) has sparked a paradigm shift in educational technology, particularly in the automation of academic assessment. While prior research has evaluated LLMs' grading capabilities across diverse formats, including multiple-choice questions [1], language teaching essays [2], [3], short open-ended responses [4], and project or course reports [5], [6], the exploration of LLMs applied to grading graduate-level research reading reports remains limited.

Developing the capacity for critical synthesis of research literature is a cornerstone of graduate education, especially in rapidly evolving fields like Software Engineering. The ability to understand and critically evaluate emerging paradigms from the latest research serves as a pivotal component of a graduate student's professional toolkit. Consequently, the reading report acts as a vital pedagogical instrument to both cultivate and evaluate these high-order cognitive skills.

However, evaluating such reports presents unique challenges. Effective assessment necessitates that the graders possess a profound understanding of selected software engineering papers and verify both the interpretive accuracy and the analytical depth of the student's work. In practice, this process is prohibitively labor-intensive. Graders, usually Teaching Assistants (TAs), must dedicate substantial time to reading and reviewing, a challenge further exacerbated by increasing class sizes. Furthermore, the reliability of grading is often compromised by the diverse backgrounds of TAs, who may lack the specific domain expertise in software engineering required to provide rigorous assessments.

In this paper, we develop an LLM-assisted workflow by modeling how humans grade research paper reports for graduate education. Using 180 real submissions from a graduate Software Engineering course, we test whether LLMs can align with human grading and explore grading inconsistency, with model and history as variables. We note that inconsistency in grading practices is seen as a primary obstacle to fairness and equity in education [7].

We find that LLMs differ substantially in their intra-model grading consistency, accompanied by remarkable inter-model inconsistency. Meanwhile, simple ensemble methods fail to enhance their alignment with human grading standards. Furthermore, we find that interaction history significantly impacts assessment fairness. Finally, we provide practices to help educators use LLMs more reliably in the grading process.

## II. Related Works

### A. LLMs in Grading Report Assessment

Recent studies have explored the potential of LLMs to assist in assessing academic reports. Matsukawa et al. [6] develop a formative assessment system to evaluate students' course reports, demonstrating how LLMs can provide structured guidance. For more complex scenarios, Chen et al. introduce CoGrader [5], a project reports assessment framework that integrates LLMs into the instructor's workflow.

However, graduate-level research paper reports require deeper domain knowledge than course or project reports, both to assess an accurate description and to evaluate the student's independent critical analysis.

### B. Factors to Influence LLMs' Grading Performance

Stahl et al. [2] investigate different prompting strategies for joint essay scoring and feedback generation, finding that the specific design of the prompt affects the quality of the model's evaluation. Molfese et al. [1] identify that the requirement of responses from LLMs may affect their performance. Kostic et al. [3] report that for the same assignment, GPT may return significantly different scores across separate attempts.

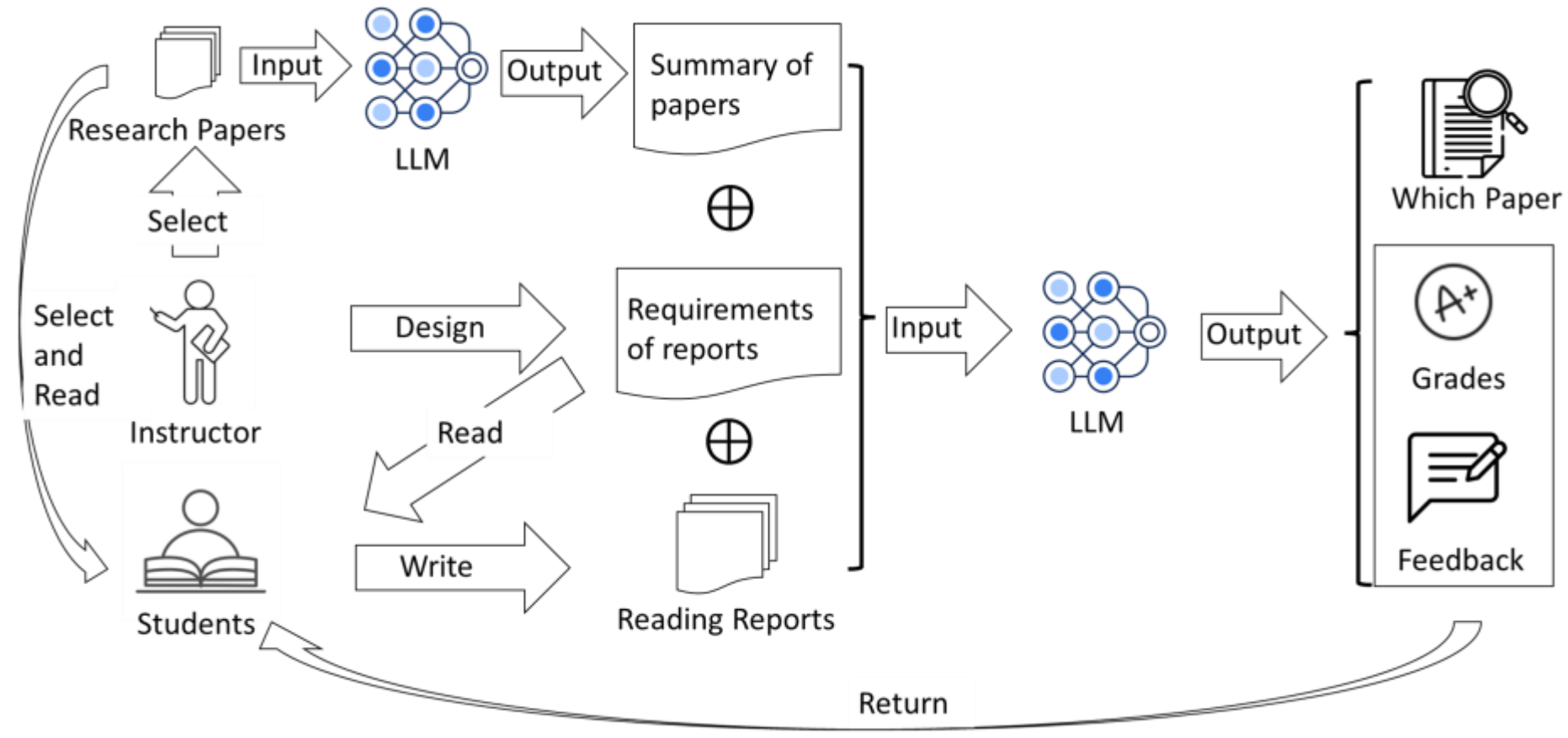

Fig. 1. The workflow to use LLM to assist in grading reading reports of research papers.

In general, LLMs could be influenced by context [8]. Context is everything the model sees at inference time, a critical component of which is the interaction history [9]. An important question arises: How does history affect its grading in reading reports?

## III. Workflow

We first outline the manual grading workflow used in the Software Engineering course for reading reports, then present the LLM-assist workflow that mimics human grading, along with the prompts for LLM integration.

### A. Original Manual Workflow

The instructor selects a set of research papers, formulates the specific requirements for the report, such as defining essential sections, word counts, and content expectations, and releases them to the students. Each student selects and reads one of the papers and composes a reading report based on the specified criteria. The TAs first familiarize themselves with all the source papers. Subsequently, they evaluate the student submissions sequentially; by referencing both the assignment rubrics and the source texts, they assign scores and provide constructive feedback. Finally, the graded reports are returned to the students.

### B. Our Proposed LLM-Assisted Workflow

Our proposed LLM-assisted workflow aligns with the manual grading design described in Section III-A. Fig. 1 illustrates our proposed workflow for utilizing an LLM to grade research paper reading reports. First, the instructor selects and inputs research papers into the LLM to generate corresponding summaries, which simulate the reading process of TAs and serve as reference material, then defines the assignment requirements and announces them to the students along with the selected papers. Subsequently, students select and read one of the selected papers and compose their reading reports in accordance with the requirements. Upon receiving the submissions, the instructor concatenates the student's report, the paper summary, and the grading requirements to feed into LLM. The model is then guided to provide the paper the reading report for, a letter grade, and feedback. The identification of papers serves as a preliminary hallucination check for the instructor. The grades and feedback are returned to students, enabling them to identify specific weaknesses for further improvement.

### C. Our LLM Integration

Since prompt engineering is not the focus of this study, our prompt was predominantly generated by Gemini Web interface; we present the skeleton of the prompt below:

**For Paper Summary**

You are an academic reading assistant. Summarize multiple papers for assignment grading
...
`[All Research Papers]`

**For Grading Reading Reports**

You are a strict but constructive teaching assistant grading student assignments
...
`[(optional) History]`
...
Please grade according to the following assignment requirements:
`[Requirements]`
...
Required JSON fields:
paper: ... grade: ... strengths: ... improvements: ...
...
Please use the following reference paper summary to grade:
`[(Generated by ``For Paper Summary'')Summary]`
...
File: `[One Reading Report Submission]`

Notes: We designed the workflow for students' submissions to be handled one at a time[1]. The [History] encompasses all previous interaction records.

## IV. Research Design and Analysis

### A. Research Questions

We aim to answer the following research questions:

RQ1 How consistent is the grading within the same model and across different models? Can ensembling make them more aligned with human grading?

RQ2 Does the grading history affect the results? Do different grading sequences lead to inconsistent scores?

### B. Study Design

Our technical evaluations are designed to assess the effectiveness and fairness of LLM grading in educational settings.

*1) Data Collection:* This study utilizes data collected from a 2025 graduate-level Software Engineering course. The research design, spanning data collection to assessment, encompasses the following key components:

- ① **Selected Papers:** The instructor curated seven research papers from diverse sub-fields of Software Engineering.
- ② **Requirements:** The instructor defines the assignment requirements.
- ③ **Which Paper to Read:** Utilizing the Canvas LMS, students selected one paper from the list.
- ④ **Student Submissions:** A total of 180 valid reading reports were collected.
- ⑤ **Human Grading:** Each submission was graded by one of five TAs, all of whom are Computer Science PhD students. Submissions that violated the provided requirements (e.g., word count or content constraints) received low scores ($\leq B$). Otherwise, grading was based on the subjective judgment of the assigned TA.

*2) Model Selection:* We adopt Grok (Grok-4.1-Fast [10] from xAI) and GPT (GPT-oss-120b [11] from OpenAI) in our study. We call their API through OpenRouter. We set the LLM temperature to 0 to make it focus [4]. These models are selected for their widespread use [2].

*3) Metrics:* To evaluate grading consistency across multiple assessments of the same set of assignments, we analyze both grades and rankings.

For grade-level consistency, we employ the Wilcoxon signed-rank test (Hypothesis: Two-sided, significance level: $\alpha = 0.05$) [12] to determine if there are significant differences between two grading rounds. As a non-parametric alternative to the paired t-test, the Wilcoxon test assesses whether the median scores of the two assessments differ significantly, remaining robust even if the grade distribution is non-normal. For ranking-level consistency, we use the Intraclass Correlation Coefficient (ICC, Model: Two-way Mixed, Type: Consistency) [13] to measure the similarity of students' relative standings. It evaluates the reliability of the relative ranking provided by the pair of assessments, focusing on whether different sets of grades maintain a consistent order of stu-dents, regardless of systematic shifts in absolute scores. The interpretation of the ICC value [13] is shown in Table I.

TABLE I
Interpretation of ICC values

| ICC Value Range | Consistency Strength |
|---|---|
| $< 0.50$ | Poor |
| $0.50 \leq \text{ICC} < 0.75$ | Moderate |
| $0.75 \leq \text{ICC} < 0.90$ | Good |
| $\geq 0.90$ | Excellent |

To further validate the alignment between LLM assessments and human grading, we analyze the hit rate of lower-tier assignments, denoted as $Hit@k$. Given the subjectivity of grading reading reports, human scores naturally exhibit variance. Since the Software Engineering courses adopt an active grade appeal system, we assume that low grades (Grade B or below, bottom 16.11%) that remain unchallenged by students are reliable indicators of lower-quality submissions. Consequently, for each LLM assessment, we measure the percentage of these human-identified low-grade submissions that fall into the bottom $k\%$ (where $k = 20, 30, 40$) of the LLM-generated rankings. The hit rate is calculated as follows:

$$Hit@k = \frac{N(\text{Human Grade} \leq B \cap \text{LLM Rank} \in \text{Bottom } k\%)}{N(\text{Human Grade} \leq B)}$$

We also perform a preliminary assessment for potential hallucinations by cross-checking the LLM-generated paper selections against the students' actual paper selections, as discussed in Section IV-D.

*4) Experimental Procedure:* We collect the data in IV-B1 produced by the Original Manual workflow stated in Section III-A. After that, these data, except for the human grading results, are used as components of our proposed LLM-Assisted Workflow (see Fig. 1) to obtain grading results.

For RQ1, we utilize all 180 student submissions, evaluated by both Grok and GPT. Each model performed the assessment three independent times without history by the same prompt, resulting in a total of six sets of grading data. We conduct pairwise Wilcoxon signed-rank tests and calculate the ICC and $Hit@k$ across these six assessments to evaluate consistency and alignment. We also calculate the average scores of the three runs for each model and evaluate the resulting $Hit@k$. For RQ2, we select a subset consisting of the first 50 student submissions, with Grok serving as the evaluation model. We establish three experimental conditions to investigate the impact of history:

- ① (Ind.) Independent grading without history.
- ② (Asc.) Sequential grading with history, in ascending alphabetical order of submission file name.
- ③ (Desc.) Sequential grading with history, in descending alphabetical order of submission file name.

Each experimental condition is executed for three independent runs. We calculate the average scores across these three runs

[1]The Gemini Web interface supports a maximum of only 10 uploads, which would also raise questions regarding interaction history.

[2]Access to other prominent models API, such as Gemini or specific closed-source GPT variants, remains restricted in the authors' region.

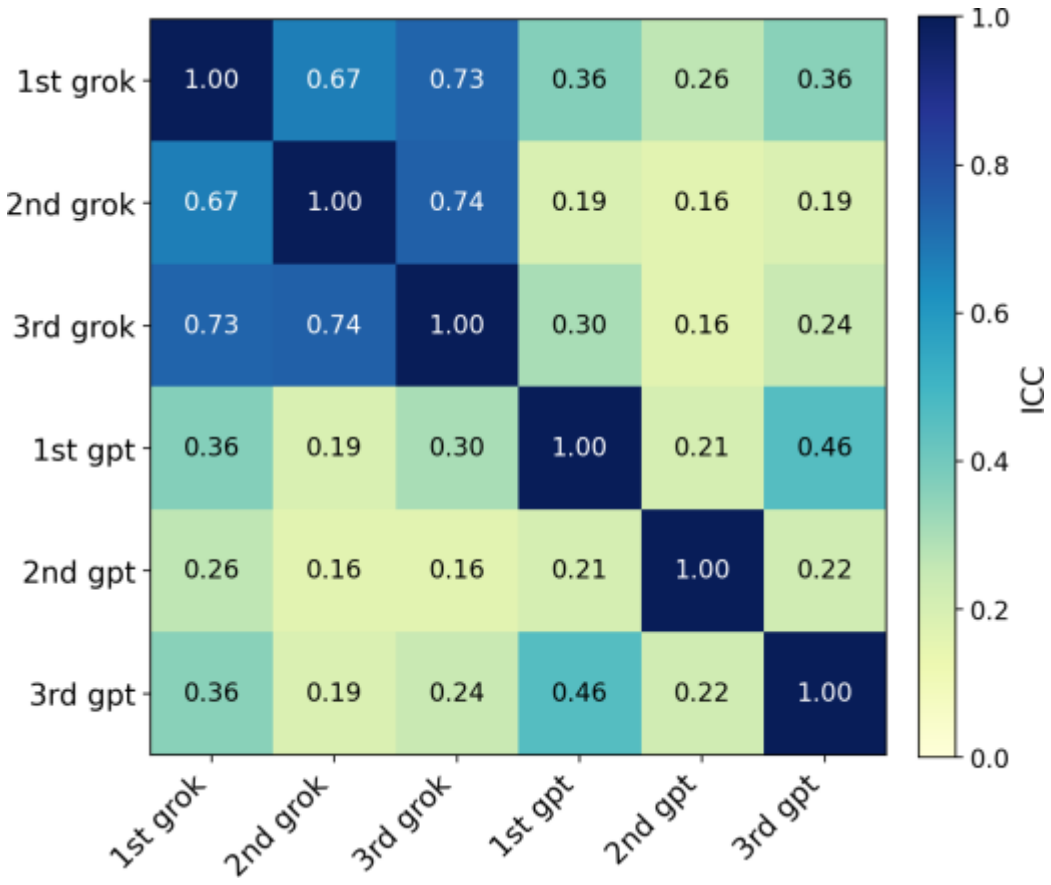

Fig. 2. Intraclass Correlation Coefficient across paired attempts and models.

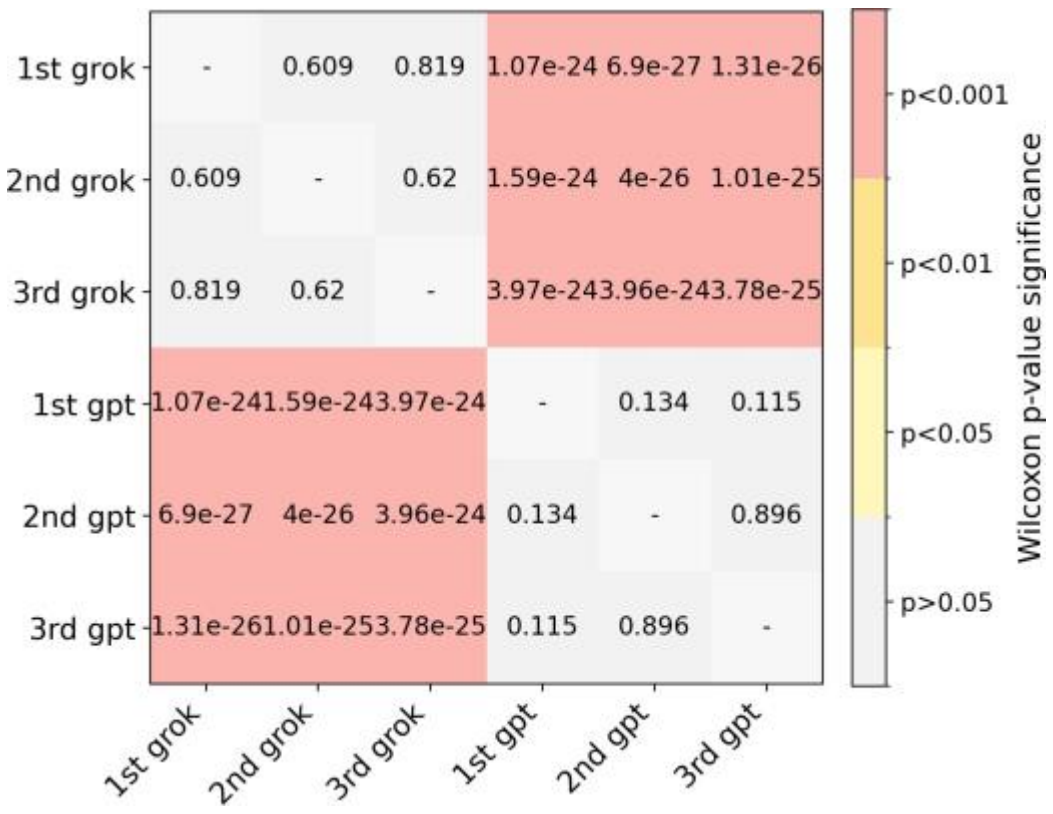

Fig. 3. P-values in the Wilcoxon signed-rank test across paired attempts and models.

before conducting the pairwise Wilcoxon signed-rank test and calculating ICC and *Hit@k*.

### C. Study Results and Analysis

*1) Answer to RQ1:* We first analyze the consistency within and across models. As illustrated in the ICC heatmap (Fig. 2), both models exhibit different levels of internal consistency. Grok maintains a relatively stable ranking across attempts ("Moderate"), whereas GPT shows a broader range of fluctuations in its scoring logic ("Poor"), aligned with findings in [3]. When comparing the two models, the ICC values also fall into "Poor", indicating that different LLM architectures apply fundamentally different internal rubrics despite being provided with the same grading requirements in the same prompt. This is further supported by the Wilcoxon signed-rank test results in Fig. 3, where the $p$-values between GPT and Grok attempts are consistently below 0.05, confirming the statistical distinctiveness.

Regarding grading accuracy, Table II presents the *Hit@k* metrics. Grok generally demonstrates a higher capability in identifying lower-tier assignments compared to GPT. For example, at $k = 40\%$, Grok's hit rate reaches up to 55.17% in its third attempt, while GPT's best performance at the same threshold is 48.28%. Interestingly, the "Avg. GPT" and "Avg. Grok's scores do not consistently outperform individual runs across all $k$ values (e.g., for $k = 20\%$, Avg. GPT achieves results lower than any of the attempts of GPT), suggesting that simple ensemble averaging may not be a final solution for eliminating LLM grading volatility.

**Recommendations for Educators**: Educators should conduct task-specific cross-model validation for model selection before deploying LLMs for student assessment, rather than only relying on simple score ensembles to eliminate inconsistencies.

TABLE II
HIT@K RESULTS FOR DIFFERENT ATTEMPTS WITH DIFFERENT MODELS

| | $k = 20\%$ | $k = 30\%$ | $k = 40\%$ |
|---|---|---|---|
| 1st GPT | 24.14% | 27.59% | 48.28% |
| 2nd GPT | 27.59% | 34.48% | 41.38% |
| 3rd GPT | 24.14% | 27.59% | 31.03% |
| Avg. GPT | 17.24% | 31.03% | 37.93% |
| 1st Grok | 24.14% | 31.03% | 48.28% |
| 2nd Grok | 27.59% | 37.93% | 51.72% |
| 3rd Grok | 31.03% | 34.48% | 55.17% |
| Avg. Grok | 24.14% | 41.38% | 51.72% |

TABLE III
WILCOXON SIGNED-RANK TEST RESULTS FOR DISTRIBUTION DIFFERENCES ACROSS GRADING MODES

| Condition 1 | Condition 2 | $p$-value | Decision on $H_0$ |
|---|---|---|---|
| ② (Asc.) | ③ (Desc.) | 0.151 | Fail to Reject |
| ② (Asc.) | ① (Ind.) | <0.001 | Reject |
| ③ (Desc.) | ① (Ind.) | <0.001 | Reject |

*Note:* $H_0$ assumes no significant difference in distributions.

*2) Answer to RQ2:* Table III shows the results of the Wilcoxon signed-rank test across these modes. We find a significant difference ($p < 0.001$) when comparing independent grading to either sequential mode (Asc. or Desc.). This indicates that the presence of a continuous chat history causes the model to shift its grading. However, the difference between Ascending and Descending orders is not statistically significant ($p = 0.151$), suggesting that while the existence of history influences the model, the specific alphabetical sequence of students is less impactful on the scores.

Ranking consistency also suffers when history is introduced. As shown in Table IV, the ICC between the Ascending and Independent modes is only 0.374 ("Poor"), whereas the consistency between the two sequential modes is 0.532 ("Moderate"), similar to that between Descending and Independent, lower than the consistency within different attempts of Grok (see Fig. 2, 0.67 at worst). This suggests that, in terms of ranking, the impact of sequence order within the history may be comparable to the impact of whether or not the history is present. Crucially, it implies that identical submissions receive different rankings based solely on their position, introducing bias that undermines equity.

Finally, Table V highlights that Independent grading achieved the highest *Hit*@20% (41.67%), compared to Ascending (33.33%) and Descending (25.00%). These results suggest that evaluating reports in isolation (without history) provides a more reliable alignment with human judgment for identifying low-quality submissions.

**Recommendations for Educators**: Recognize that grading history and submission sequences can compromise fairness if using LLM for grading. Initiate fresh sessions for each batch of submissions until effective mitigation strategies are developed.

TABLE IV
INTRACLASS CORRELATION COEFFICIENT (ICC) ANALYSIS ACROSS GRADING MODES

| Condition 1 | Condition 2 | ICC | Interpretation |
|---|---|---|---|
| ② (Asc.) | ③ (Desc.) | 0.532 | Mod. |
| ② (Asc.) | ① (Ind.) | 0.374 | Poor |
| ③ (Desc.) | ① (Ind.) | 0.511 | Mod. |

TABLE V
HIT@K RESULTS ACROSS GRADING MODES

| | $k$ = 20% | $k$ = 30% | $k$ = 40% |
|---|---|---|---|
| Desc. | 25.00% | 33.33% | 50.00% |
| Asc. | 33.33% | 33.33% | 33.33% |
| Ind. | 41.67% | 41.67% | 50.00% |

### *D. Discussion*

**Why and how does history matter?** Recent studies show that LLMs still struggle with understanding long context [14]; different histories, as part of context, may divert the attention of LLM from the student's submission. Unfortunately, our further exploration finds no simple linear relation between revised count, mean score, and score variance, suggesting the issue is non-trivial.

**Are hallucinations present?** In our experiments for RQ1 and RQ2, the papers identified by LLMs from reading reports are all consistent with the students' actual selections. Our manual spot-check of Grok's results in RQ1 finds no factual errors in the generated feedback. Hallucination may no longer be a serious issue for such tasks in the latest LLMs.

**Threats to Validity.** Because the study relies on APIs, model outputs are occasionally unstable, truncated, or missing. We therefore only compare valid responses, resulting in an effective sample size smaller than the original. We conducted experiments using Python scripts, which may contain bugs despite efforts to test and fix them.

## V. CONCLUSION

This paper introduces an LLM-assisted grading workflow for research paper reports, demonstrating its potential to re-duce the pedagogical burden of graduate-level assessment for educators. Our empirical analysis reveals that model selection and interaction history are critical determinants of grading reliability. To improve fairness and effectiveness, we recommend that educators perform task-specific validation for LLM model selection and evaluate submissions in independent sessions to mitigate history-induced bias. Nevertheless, we caution that complete reliance on LLM grading may introduce systemic inequities. Future work will expand the scale of experiments and explore additional factors, such as gender, to examine whether they result in biased outcomes. We will also evaluate mitigation strategies, including prompting LLMs to ignore historical information. Finally, we aim to develop AI agents that repurpose biased grading histories as constructive resources to support fair assessment in graduate education.

## ACKNOWLEDGMENT

We thank the anonymous reviewers for insightful comments.